\documentclass[preprint]{elsarticle} 

\usepackage{times}
\usepackage{amsmath}
\usepackage[]{graphicx}
\usepackage{algorithm}
\usepackage{algorithmic}
\usepackage{verbatim} 
\usepackage[color]{changebar}

\makeatletter
\def\ps@pprintTitle{%
 \let\@oddhead\@empty
 \let\@evenhead\@empty
 \def\@oddfoot{\centerline{\thepage}}%
 \let\@evenfoot\@oddfoot}
\makeatother

\begin{document}

\begin{frontmatter}

\title{Optimized Secure Position Sharing \\ with Non-trusted Servers} 

\author{Pavel Skvortsov, Bj\"orn Schembera, Frank D\"urr, Kurt Rothermel}
\address{Institute of Parallel \& Distributed Systems (IPVS), Universit\"at Stuttgart \\
Universit\"atsstrasse 38, 70569 Stuttgart, Germany \\ Tel.: +4971168565802; Fax: +4971168565832}
\ead{skvortsov@hlrs.de, schembera@hlrs.de, frank.duerr@ipvs.uni-stuttgart.de, kurt.rothermel@ipvs.uni-stuttgart.de}

\begin{abstract}

\noindent
Today, location-based applications and services such as friend finders and geo-social networks are very popular.
However, storing private position information
on third-party location servers leads to privacy problems. 
In our previous work, we proposed a position sharing approach for secure management of positions on non-trusted
servers \cite{Duerr2011,Skvortsov2012}, which distributes 
position shares of limited precision among servers of \emph{several} providers. 
In this paper, we propose two novel contributions to improve the original approach.
First, we optimize the placement of shares among servers by taking
their trustworthiness into account. 
Second, we optimize the location update protocols to minimize the number of messages
between mobile device and location servers. 



\end{abstract}


\end{frontmatter}

\section{Introduction}

\noindent
Driven by the availability of positioning systems such as GPS, powerful smartphones such as the iPhone or Google
Android phones, and cheap flat rates for mobile devices, location-based services enjoy growing popularity.
Advanced location-based applications (LBAs) such as friend finders or geo-social networks
are typically based on location server (LS) infrastructures that store mobile user positions to ensure scalability
by enabling the sharing of user positions between multiple applications. This principle relieves the mobile device
from sending position information to each LBA individually. Instead, the mobile device updates the position
at the LS, and LBAs query the LS for the positions of mobile objects.

Although certainly useful from a technical point of view, storing data on LSs raises privacy concerns if
the LSs are non-trusted. Multiple incidents in the past have shown that
even if the provider of the LS is not misusing the data, private information can be
revealed due to attacks, leaking, or loss of data \cite{privacybreaches,Mokbel2007,Pedreschi2008}.
Therefore, the assumption of a trusted LS is questionable, and technical concepts are needed
to ensure location privacy without requiring a trusted third party (TTP). 

Various location privacy approaches have been proposed in the literature. Many approaches such as $k$-anonymity
(e.g., \cite{Kalnis2007}), rely on a TTP and are, therefore, not applicable to non-trusted environments. Approaches without
the need for a TTP mainly rely on the concept of spatial obfuscation, i.e., they reduce the precision of
position information stored on the LS. However, this severely impacts the quality of service of LBAs
since they can only be provided with coarse-grained positions, depending on the degree of obfuscation. 

To solve this conflict between privacy and quality of service, we have proposed the concept of position sharing in our
previous work \cite{Duerr2011,Skvortsov2012}. 
The basic idea of this concept is to split up the precise user position into position shares of limited precision.
These shares are distributed among multiple LSs such that each LS has a limited, coarse-grained
view onto the user position. LBAs are provided with access rights to a certain number of shares on different LSs. By using share
fusion algorithms, a position of higher precision can be calculated. Therefore, LBAs can be provided with different,
individual precision levels (privacy levels) although each LS only manage less precise information. This approach does
not require a TTP. Furthermore, it provides graceful degradation of privacy, where no LS is a single point
of failure with regard to privacy. Instead, the precision of positions revealed to an attacker increases with
the number of compromised LSs.

The concept of position sharing can be applied to various settings and use cases.
For instance, it can be used to implement a secure personal data vault for location information
of individual users over a non-trusted server infrastructure. The concept of a personal data vault has been
first proposed in \cite{MunKSEHG14}. A data vault is a data repository controlled by the user for storing personal
data and controlling the accessed to data from different services. With the advent of cloud computing, it seems
to be attractive to implement data vaults atop cloud computing infrastructures, relieving the user from operating
her own dedicated servers. However, the providers of these cloud infrastructures might be non-trusted. Position
sharing offers the possibility to avoid storing all location data of the data vault at a single provider.
Instead, the data vault can be distributed among servers of different cloud providers, where each provider
only has a well-defined limited view onto the personal location information.

As another use case for position sharing, consider a startup company that wants to provide location-based
services to its customers. However, as typical for many startups, the startup company does not own a dedicated
server infrastructure but instead utilizes an infrastructure as a service (IaaS) of a third-party IaaS provider.
Assume that the customer trusts the startup company to handle his private location information securely. However,
although the customer trusts the startup company, this does not imply that the startup company trusts the IaaS
provider operating the physical servers. Position sharing enables the startup company to distribute its valuable
private customer data to several third-party IaaS providers to avoid a single point of failure and provide a
trustworthy virtual service to its customers over non-trusted IaaS infrastructures. 

In this work, we present two novel contributions to extend the original position sharing approach:
(1) an algorithm for optimizing the placement of position shares on location servers of \emph{different}
trustworthiness; (2) optimized share update protocols that significantly reduce the communication
overhead of the original approach.

In our previous approach, we made the simplified assumption that every LS provider is
equally trustworthy, i.e., each LS has the same risk of being compromised.
However, a user might trust certain providers such as big companies with good reputation more
than others. Therefore, it seems reasonable that LSs of providers of higher
trust levels should store more precise information, i.e.,
either position shares of higher precision or more shares.
For instance, in the examples above IaaS providers operated by companies like Google, Microsoft
or Amazon might be considered more trustworthy than servers of a cheaper but not so well-known
IaaS providers. Still it might be reasonable to utilize such cheaper providers for
monetary reasons. With our extended position sharing approach, we can balance the risk
of revealing data by considering the individual trustworthiness of providers.

Therefore, the first contribution of this paper is an improved position sharing approach that takes individual trust levels
of LS providers into account. To this end, we optimize the placement of shares on LSs
to increase the protection of privacy. We propose a suitable privacy metric and scalable share placement
algorithms that (a) flexibly select $n$ of LSs, and (b) balance the risk
among providers such that the risk of disclosing private information stays
below a user-defined threshold. Moreover, we aim to meet the user-defined privacy requirements with a
minimum number of LSs to minimize the overhead of updating and querying several LSs.

Moreover, in our previous work we did not consider 
how multiple position updates affect the communication overhead. 
This factor can negatively affect the scalability of our approach, since each update includes $n$ messages (where $n$
is the number of LSs). Therefore, as the second contribution
of this paper, we propose a position update algorithm which improves scalability by minimizing the number of transmitted messages
and does not change the user's desired location privacy levels. This improvement is achieved by omitting updates of shares
that can remain unchanged after the given position change.
Our evaluations show that the proposed method can save up to 60\% of messages 
compared to the previous version of this approach.

The rest of this paper is structured as follows: In Section 2, we give an overview of the related work. In Section 3,
we describe our system model and privacy metric. In Section 4, we present the basic position sharing approach including the share fusion and
share generation algorithms. In Section 5, we propose a share placement algorithm which distributes position shares among LSs
depending on their individual trustworthiness. Then in Section 6, we present position update algorithm which optimizes
the communication overhead caused by multiple position updates. Finally, in Section 7, we conclude this paper with a summary.

\section{Related Work}

\noindent
In this section, we will discuss existing approaches for location privacy. For a more in-depth analysis
and classification of location privacy techniques, we refer to our survey paper \cite{WSDKpuc2012}.

A classic solution employed to preserve location privacy is \emph{cryptography}.
However, if user positions stored on servers are encrypted, server-side query
processing of advanced queries such as range queries over the encrypted
data becomes impossible, or it is possible only at higher cost \cite{Riboni08}. 


Another example of a cryptography-based approach for location privacy was proposed by
Mascetti et al. \cite{Mascetti2011} to implement proximity services for geo-social networks. The authors assume
that service providers are non-trusted and consider the scenario where mobile users want
to notify their friends called buddies in their proximity. The main idea is that the secret keys
are shared with the selected buddies in a distributed fashion and remain unknown to the
service providers. The authors use a precision metric which is defined through the union of
multiple discrete space cells called granules. A drawback of this approach is that it requires a
complex implementation of the encryption functionalities. Similarly to the work of Zhong et al. \cite{zhong2007},
this approach only considers specific friend-finder and proximity calculation scenarios.

Another method to preserve location privacy is to send \emph{dummy positions} to LBAs together with the actual user position \cite{Kido2005}.
The problem with this approach is, however, that dummy positions can be easily distinguished 
if the attacker has some background information such as database of real user movements \cite{ShankarGI09}.

The idea of \emph{mix zones} \cite{BeresfordS04} is to select privacy-sensitive areas called mix zones in which users
do not send position updates while they are visible outside the mix zones. Some extensions try to avoid the threat of
analyzing possible user trajectories based on the known entry and exit points on the borders of a mix zone \cite{Palanisamy2011}.
The mix zones approach lacks flexibility since it needs a pre-defined division of space
into zones, and does not allow for various privacy levels in different zones.

Many existing location privacy approaches are based on $k$-\emph{anonymity}, e.g., \cite{mokbel06}. The main idea
is to send a set of $k$ different positions of real mobile users ($k$-set) to the LBA, such that the actual user position
is indistinguishable from $k-1$ other positions.
Many extensions of $k$-anonymity aim to make the $k$-set more robust to various attacks---mostly
against attacks based on the analysis of user attributes, e.g., \cite{Machan07,Bamba08,Solanas2008,Li2007,WongLFW06}.
However, in order to select a $k$-set, a trusted anonymizer with global
view is needed, which requires trust to a TTP, introducing a single point of failure.

\emph{Obfuscation} approaches such as \cite{Ardagna07} deliberately decrease the precision of user positions stored on servers.
A TTP is not required in this case since the cloaked region can be generated by the user independently,
but the queries over obfuscated locations can result in imprecise or probabilistic answers.
Our approach is also based on spatial obfuscation, but it gracefully degrades position precision depending on
the number of missing position shares, and it supports multiple obfuscation levels (i.e., privacy levels).

Marias et al.\ \cite{Marias2005} propose to apply the concept of \emph{secret sharing} \cite{Shamir79} to position
information to distribute the information about single positions among several servers.
Their secret sharing approach relies on cryptographic techniques, which means that all shares are required in order to retrieve a position.
In order to overcome this drawback, we proposed the \emph{position sharing} approach based on spatial
obfuscation \cite{Duerr2011}, map-aware spatial obfuscation \cite{Skvortsov2012},
and cryptographic multi-secret sharing techniques (PShare) \cite{Wern1203:PShare}.
However, in all these works, LSs are assumed to be equally trustworthy, and, therefore, each LS stores one
position share of equal precision. In this paper, we will show that optimized share placement
based on the individual trust levels of providers significantly increases privacy if providers
are of different trustworthiness and opens up the possibility to minimize the number
of required LSs. Moreover, we will extend our previous work by introducing an optimized location
update algorithm, which reduces the number of transmitted messages to the LSs.

\section{System Model and Privacy Metric}
\label{sec:sys_mod}

\noindent
This section introduces our conceptual system model, operational system model and privacy metrics of the position sharing approach.

\subsection{System Model}

\begin{figure}
  \begin{center}
    \includegraphics[width=0.65\textwidth]{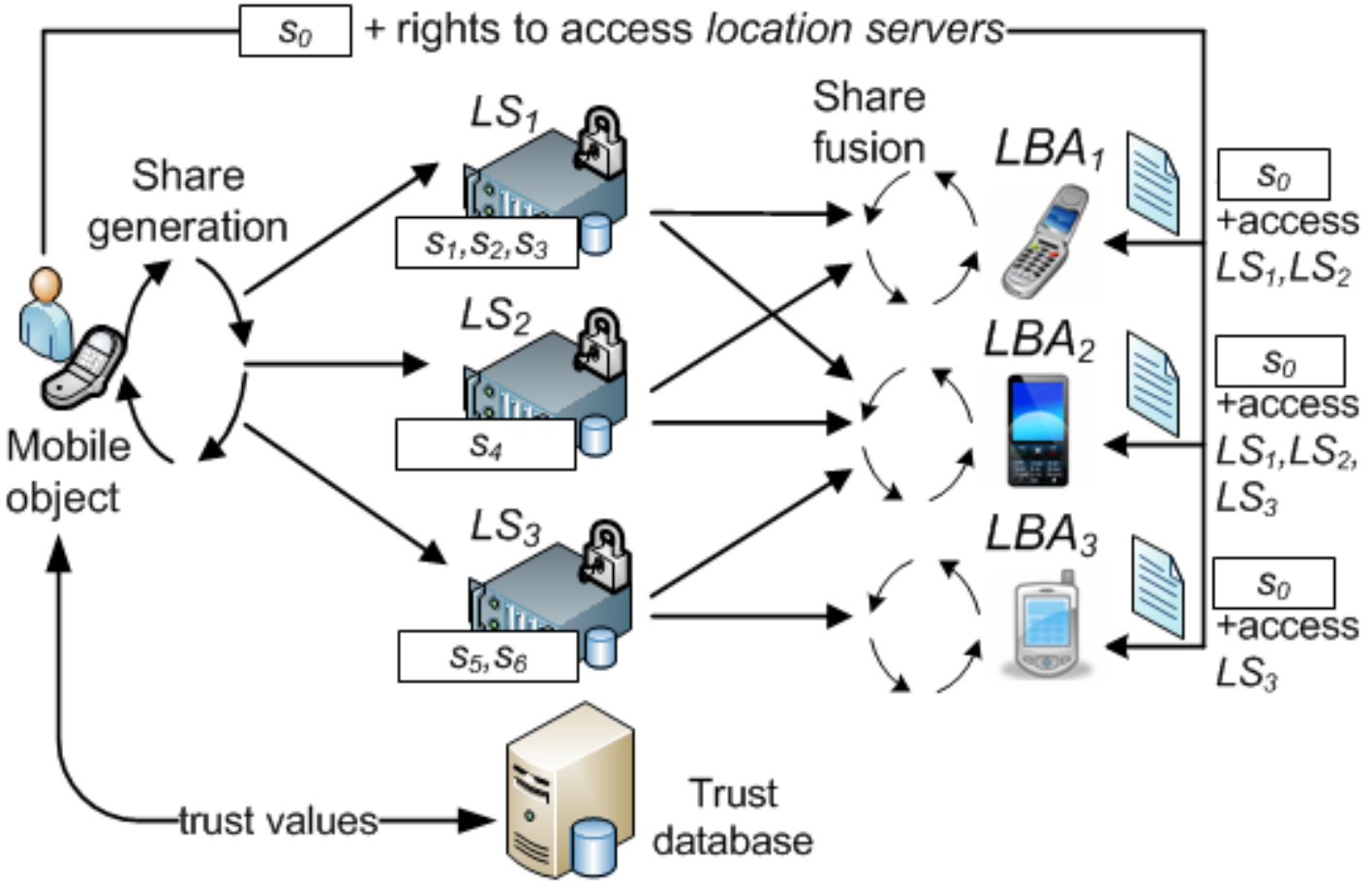}
    \caption{System model}
    \label{fig:system_model}
  \end{center}
\end{figure}

\noindent
Our system model is shown in Figure~\ref{fig:system_model}. It consists of four components:
\emph{mobile objects}, \emph{location servers}, \emph{location-based applications}, and a \emph{trust database}.

The user is represented by the \emph{mobile object (MO)}, 
which knows its precise position $\pi$, for instance, determined through GPS. 
A position of certain \emph{precision} is defined by a circular area which we call \emph{obfuscation area},
where radius $r$ of this circular area defines the
\emph{precision} $\mathrm{prec}(\pi) = \phi = r$ of position $\pi$. A smaller radius corresponds to a higher precision:
if $r_1 = \mathrm{prec}(\pi_1), r_2 = \mathrm{prec}(\pi_2)$ and $r_1 < r_2$,
then the precision of $\pi_1$ is higher than the precision of $\pi_2$.

The MO executes a local component to perform the generation of position shares on the mobile device.
We assume that this component can be implemented in a trustworthy way, e.g., by using a Trusted Computing Platform \cite{TCP07}.
The MO generates one \emph{master share} $s_\mathrm{0}$ with the minimal acceptable precision $\phi_{min}$
chosen such that there are no problems with regard to privacy, and a set of $n$ refinement shares $S = \{s_1,\ldots,s_n\}$:
\begin{equation}
\mathrm{generate}(\pi, n, \phi_{min}) = \{s_0, S\}
\end{equation}
Given a subset $S' \subseteq S$ of refinement shares, its fusion with $s_{0}$ (which is known to everyone) results
in a position $\pi'$ of a certain well-defined precision:
\begin{equation}
\label{fuseeq}
\mathrm{fuse}(s_{0}, S') = \pi', \\
\end{equation}
\begin{equation*} 
\textrm{~with~} \mathrm{prec}(\pi) < \mathrm{prec}(\pi') \textrm{,~~i.e.,~~} \phi \leq \phi'
\end{equation*}
The fusion of $s_{0}$ with the set $S$ of all
refinement shares obtained from the LSs provides the exact position $\pi$ of precision $\phi_{max}$.

We say that shares are \emph{heterogeneous} if 
each share $s_i$ increases the position precision by an individual amount $\Delta^\phi_i$.
Typically, for heterogeneous shares, share fusion has to be performed in a certain fixed order,
in contrast to \emph{homogeneous} shares that can be fused in any order.
If a share generation algorithm produces homogeneous shares, only the number of different
shares defines the resulting precision level $\phi_k$. 
In this case, the precision increase is equal for
each share: $\Delta^\phi_1 = \Delta^\phi_2 = \ldots = \Delta^\phi_n = \phi_{max}/n$. 


A \emph{location server (LS)} stores and delivers location data of users to LBAs. Each LS
has a standard authentication mechanism, which allows to specify access rights
for the LBAs (given by a user) to access shares stored by this LS. The maximal allowed precision
of a user position is defined individually depending on the concrete LBA by specifying
a certain set of LSs accessible for each LBA.

\begin{figure}
  \begin{center}
    \includegraphics[width=1.0\textwidth]{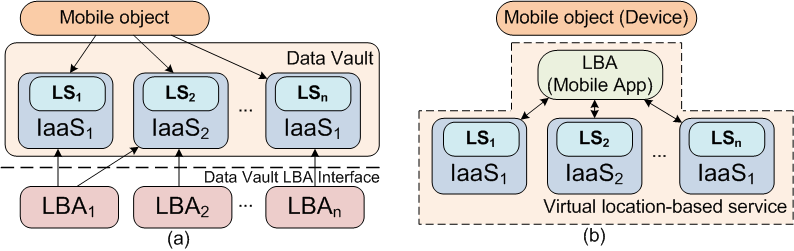}
    \caption{(a) Distributed Location Data Vault scenario;
		(b) Virtual location-based service provider scenario}
    \label{fig:opsysmod}
  \end{center}
\end{figure}

We assume that each provider operates one LS. Internally, this LS can be implemented
by a number of physical servers, e.g., running in a data center.
Also, an LS can be implemented based on the ``virtual provider'' model on top of an Infrastructure-as-a-Service (IaaS).
For instance, to implement a cloud-based personal Data Vault storing and filtering locations of individual users as already
motivated in Section 1, each LS could be implemented atop a virtual machine operated by an individual IaaS provider as depicted
in Figure~\ref{fig:opsysmod}a. The set of LS then implements the distributed personal Data Vault offering a well-defined interface to LBAs.
The LBAs need to be aware of the interface of the distributed Data Vault in order to query data from the Data Vault,
which is a typical assumption of the Data Vault concept.

Another operational model is the implementation of a virtual LBA offered by a startup company (LBA provider)
not owning a dedicated server infrastructure as motivated in Section 1. Here, we assume that the customer actually
trusts the LBA provider, however, the LBA provider does not trust the IaaS providers operating the virtual machines
running the LS as depicted in Figure~\ref{fig:opsysmod}b. Although in this setting there is only a single LBA which is trustworthy from
the point of view of the mobile object (customer) and thus can be provided with position information of maximum
precision, the position sharing concept is still useful since it allows the trusted LBA provider to implement
its service atop a non-trusted IaaS infrastructure to manage positions of the whole population of all customers.
The LBA logic is implemented by a trusted app running on the mobile device, so the non-trusted IaaS
providers cannot influence how data is distributed among LS.

Since the Data Vault scenario is the more general case with several non-trusted LBAs, which should be
provided with location information of different precision, we will further on focus on this scenario only.    


Each LS is non-trusted and can be compromised with probability $p_i$. 
Risk value $p_i$ represents the probability of $\mathrm{LS}_i$ to behave maliciously, i.e., to misuse
the user's private position information, or to be compromised by an external attacker.
The concrete 
concepts for calculating $p_i$ are beyond the scope of this paper. For instance, we can use
the \emph{generic probabilistic trust model} described in \cite{KinatederBR05}.
This model is generic in the sense that it allows mapping of various representations of trust values to the probabilistic interval $[0;1]$.
Different LSs might have different risks depending, for instance, on the reputation of their
provider. Moreover, different users might have individual trust in the same LS (and/or its provider).

The \emph{trust database} manages the trust in different LSs by providing the
probabilities $p_i$ that LS$_i$ can be compromised.
Based on the obtained risk values, the user can determine the number and set
of LSs needed to satisfy his security requirements, as we will show in the following sections. 
We assume that the trust database is given, and it
is filled with data, for example, by analyzing the feedback of other users through a
\emph{reputation system} \cite{GutscherHS08,GutscherA09}.

\emph{Location-based applications (LBAs)} query or track MO's position
and obtain multiple shares from different LSs depending on the access rights given by user.
Then, the LBA fuses the obtained shares by using
function $\mathrm{fuse}(...)$ (Equation~\ref{fuseeq}) in order to get the user position with a certain level of precision.

\subsection{Privacy Metric}

\noindent
The user's privacy levels are primarily defined through \emph{precision} levels $\phi_k$,
which are pre-defined by the user for each $0 \leq k \leq n$ as radii $r_k$ of a \emph{circular obfuscation areas}.
Additionally, we use a probabilistic privacy metric since the precision of a position obtained by an attacker depends on the
probabilities of compromising LSs as well as on the ability of an attacker to derive higher position
precisions by analyzing the obtained shares (as shown in \cite{Duerr2011,Skvortsov2012}).
The following distribution $P_\mathrm{k,attack}$ defines the \emph{probability} of an attacker obtaining a position
$\pi_\mathrm{k,attack}$ of a certain precision
$\phi_\mathrm{k,attack} = \mathrm{prec}(\pi_\mathrm{k,attack})$
depending on the number $k$ of compromised LSs: 
\begin{equation}
\mathrm{P_\mathrm{k,attack}}(\phi_k) = \mathrm{Pr}[\phi_\mathrm{k,attack} \leq \phi_k]
\end{equation}
This metric can be used by the MO to define the acceptable \emph{probabilistic guarantees} represented
as a set of probability thresholds $P_{k}(\phi_k)$ corresponding to various precision levels $\phi_k$.
For example, an MO can specify that an attacker must not be able to 
obtain a position of precision $\phi_1 \leq 1$ km with probability $P_\mathrm{1,attack} > 0.2$,
and $\phi_2 \leq 2$ km with $P_\mathrm{2,attack} > 0.1$, etc.

\section{Basic Position Sharing Approach}

\noindent
In this section, we will present the basic principle and two basic versions of the position sharing approach: (a) for
open space models (with no map knowledge) and (b) for spatial constrains (taking into account map knowledge
as explained later). Each of them includes an algorithm for LBAs to fuse position shares,
and an algorithm for MOs to generate the shares.

\begin{figure}
  \begin{center}
    \includegraphics[width=0.3\textwidth]{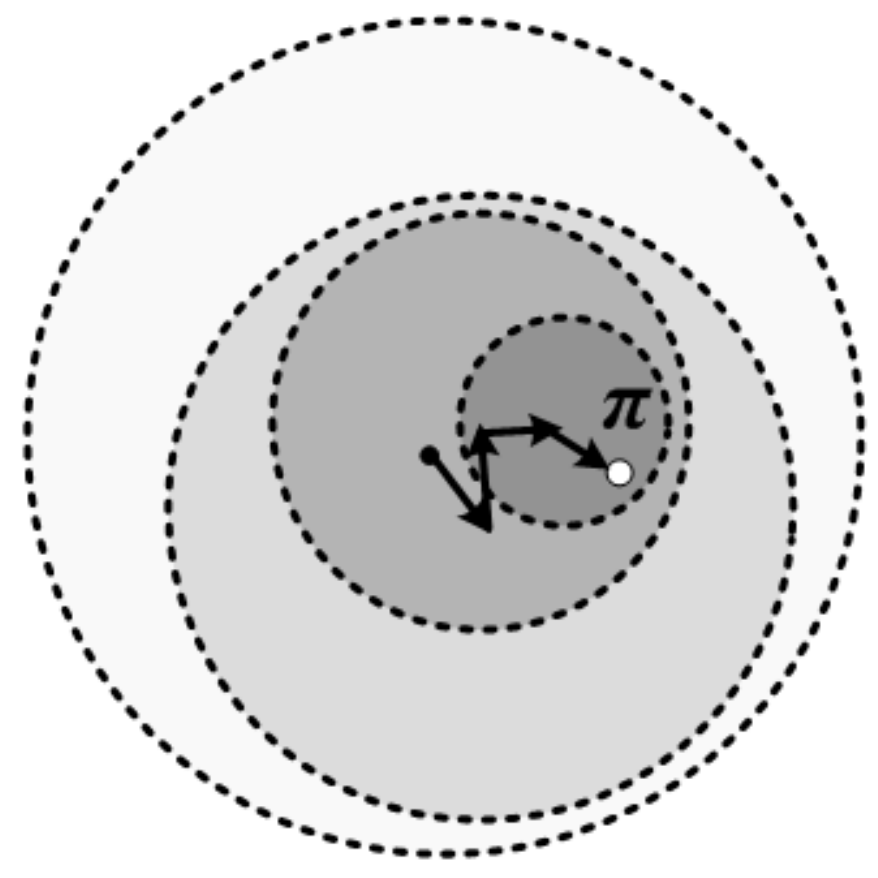}
    \caption{Basic position sharing approach: after getting each new share (i.e., a shift vector plus a radius decrease), the
    precision is increased until we get the exact target position $\pi$}
    \label{fig:position_sharing}
  \end{center}
\end{figure}

\subsection{Position Sharing: Basic Principle}

\noindent
In general, our position sharing approach is based on geometric transformations, where
imprecise geometric positions are defined by circular areas $c_i$ with radii $r_i$ \cite{Duerr2011,Skvortsov2012}. Each share is a vector
shifting the center of the current obfuscated position represented as a circle, whose radius is decreased after
every shift. An example of precision increase through such share fusion is shown in Figure~\ref{fig:position_sharing}.
The share defines the precision increase $\Delta^\phi_i$ after the corresponding $i$th
shift of the center of the circle and the radius decrease. 

After generating the shares, the MO distributes them among multiple LSs and updates them continuously.
The master share is publicly available to anyone. In order to control the access to refinement
shares, the MO defines access rights to a certain subset of shares for each individual LBA.

The concrete share generation and share fusion algorithms depend on whether we consider
the availability of map knowledge or not. Next, we first assume that no map knowledge is available
before we present an extended approach taking map knowledge into account.

\subsection{Open Space Position Sharing (OSPS)}

\noindent
In \cite{Duerr2011}, we presented the position sharing approach for open space,
referred to as OSPS (Open Space Position Sharing). The open space model 
weakens the attacker model by assuming that no map is known to an adversary. In other words,
we assume a uniform probability distribution of MO positions.

\begin{algorithm}
\small
\caption{OSPS: fusion of shares}
\label{alg1x}
\begin{algorithmic}[1]
\STATE \textbf{function} $fuse\_k\_shares\_OSPS(n, s_0, \vec{s_1} \ldots \vec{s_k})$
\STATE $\Delta r \gets r_0/n$;
$\vec{p} \gets \vec{p_0}$; 
$r \gets r_0$
\STATE \textbf{for} $i = 1$ \textbf{to} $k$ \textbf{do}
\STATE ~~~~~~$\vec{p} \gets \vec{p} + \vec{s_i};$
\STATE ~~~~~~$r \gets r - \Delta r$
\STATE \textbf{return} $c_k = \{\vec{p}, r\}$
\end{algorithmic}
\normalsize
\end{algorithm}

\noindent
The share fusion algorithm is executed on the LBA, which knows the number of LS providers
resp.~the total number of refinement shares $n$, the 
obtained refinement shares $\vec{s_1} \ldots \vec{s_k}$ ($k < n$), and the master share $s_0$.
We define the master share as the obfuscation circle $c_0$ with center $p_0$ and
precision $\phi$ defined through radius $r_0$.

The refinement shares are shift vectors $S = \{\vec{s_{1}} \ldots \vec{s_{n}} \}$.
The $i$th share defines the precision increase $\Delta^\phi_i$ after shifting
the corresponding $i$th circle and decreasing its radius.

\begin{figure}
  \begin{center}
    \includegraphics[width=0.65\textwidth]{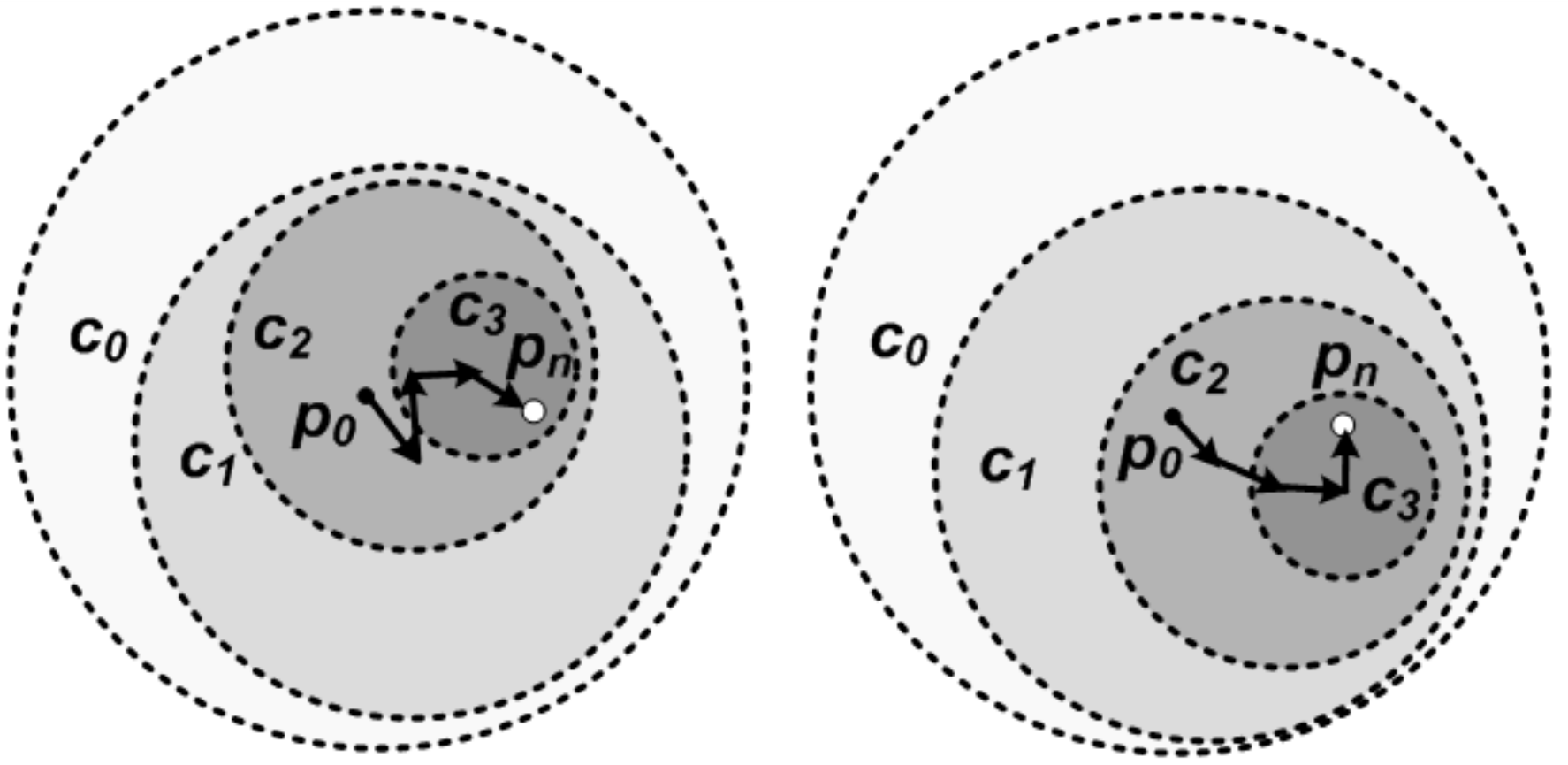}
    \caption{OSPS: same set of shares fused in an arbitrary order
		($n = 4, k = 3, r_0 = 20$ km, $\Delta r = 5$ km)}
    \label{fig:arb_ord}
  \end{center}
\end{figure}

The share fusion algorithm of OSPS is shown in Algorithm~\ref{alg1x} (cf. Figure~\ref{fig:arb_ord}).
Starting from the initial obfuscation circle $c_0$ (lines~2,5),
step-by-step for $k$ shares (line~3) each vector $\vec{s_i}$ shifts the center $p_i$ of the current obfuscation
circle $c_i$ (line~4) while reducing the radius $r_i$ (line~5) of the current obfuscation circle by a pre-defined value
$\Delta r = r_0/n = \Delta^\phi_i$ (line~2). Note that in OSPS, $\Delta^\phi_i$ has the same value $\Delta_\phi$ for each $i$.
The resulting obfuscation circle is $c_k$ (line~6).

In Figure~\ref{fig:arb_ord}, we show an example of share fusion for $n = 4, k = 3, r_0 = 20$ km, $\Delta r = 5$ km.
As shown, the order of fusing the refinement shares can be arbitrary,
while the precision (radius) of every obfuscation circle $c_k$ is well defined. 
Note that our algorithm is not dependent on the absolute values of $r_0$ and $\Delta r$, i.e.,
it works for any selected size of $r_0$.

\begin{algorithm}
\small
\caption{OSPS: generation of shares}
\label{alg2x}
\begin{algorithmic}[1]
\STATE \textbf{function} $generate\_n\_shares\_OSPS(s_0, n, \pi)$ 
\STATE $\Delta r \gets r_0/n$
\STATE \textbf{select randomly} $p_0$ \textbf{such that} $distance(\vec{p_0}, \pi) \leq r_0$
\STATE \textbf{do}
\STATE ~~~~~~\textbf{for} $i = 1$ \textbf{to} $n-1$ \textbf{do}
\STATE ~~~~~~~~~~~~\textbf{select rand.} $\vec{s_i}$ \textbf{with} $\left|\vec{s_i}\right| \leq \Delta r$ \textbf{such that} $\pi \in c_i$
\STATE \textbf{while} $distance(\vec{p_0} + \sum_{i=1}^{n-1}\vec{s_i}, \pi) > \Delta r$
\STATE $\vec{s_n} \gets \pi - (\vec{p_0} + \sum_{i=1}^{n-1}\vec{s_i})$
\STATE \textbf{return} $\vec{s_0} \ldots \vec{s_n}$
\end{algorithmic}
\normalsize
\end{algorithm}

The generation of shares in OSPS is presented in Algorithm~\ref{alg2x}:
The input parameters are the user-defined radius $r_0$ of the initial obfuscation circle $c_0$,
the total number of shares $n$, and the precise user position $\pi = p_n$.
First, we determine the maximum shift length $\Delta r = \Delta_\phi = r_0/n$ (line~2).
In the second step, position $p_0$ of the initial circle $c_0$ is selected randomly according to a uniform
distribution such that $\pi=p_n$ is inside $c_0$ (line~3). The set of the refinement shift vectors
$S = \{\vec{s_{1}} \ldots \vec{s_{n-1}} \} $ is generated with random direction and length (lines~4-6), starting from the center of $c_0$.
All shift vectors of $S$ are concatenated such that the resulting point $\pi = p_n$ 
(with $r_{n} = 0$ correspondingly) coincides with the user's position $\pi$ within $c_0$ (line~8).
The vector lengths are selected from the interval $[0; \Delta r]$ uniformly at random. Finally, MO sends the position information to
$n$ LSs, including the master share $s_0$, the radius decrease after every shift $\Delta r$
(in OSPS, $\Delta r$ is constant for all shifts), and one share $\vec{s_i}$ for each LS.
The master share is stored on every LS, while each LS stores only one refinement share.

\subsection{Constrained Space Position Sharing (CSPS)}

\noindent
The obvious limitation of OSPS 
is that it is not applicable for constrained space models.
If an attacker has map knowledge, the attacker is able to reduce the size of the generated
obfuscation areas. By excluding areas where the user cannot be located such as water surfaces, ravines, and agriculture fields,
it is possible for the attacker to intersect an obfuscation circle $c_k$ with
areas such as roads, squares, buildings, etc.
Since the total area of possible user locations in $c_k$ is smaller than the total area of $c_k$, the target
privacy guarantees---based on the size of obfuscation circles, i.e., precision---are not preserved.
To resolve these problems, we present an improved version of the position sharing approach next, which was
first introduced in \cite{Skvortsov2012}.

\begin{figure}
  \begin{center}
    \includegraphics[width=0.65\textwidth]{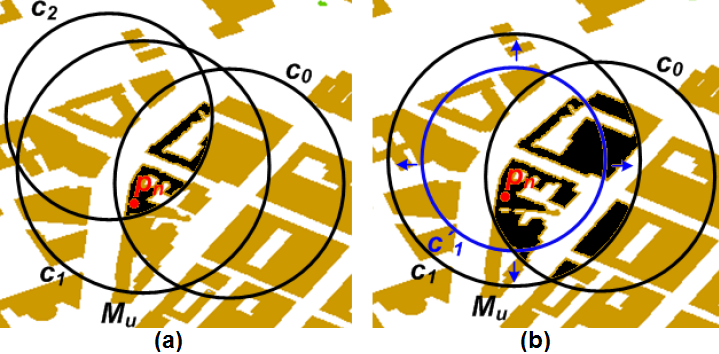}
    \caption{CSPS: a) intersection of 3 circles $c_0, c_1, c_2$ and the map representation $M_u$ (yellow);
    			b) adjustment of intersection area through radius increase for $c_1$:
    			$A_1=\mathrm{area}(M_u \cap c_0 \cap c_{1}) = \mathrm{area}(c`_{1})$}
    \label{fig:rad_inc}
  \end{center}
\end{figure}

In the map-aware position sharing approach for constrained space (CSPS), we introduce
the binary map representation $M_u$ (cf. Figure~\ref{fig:rad_inc}a; $M_u$ shown in yellow). Each location is marked by ``true'' if the MO
can be possibly located there, or ``false'' if it is impossible that MO is located in this area.
We define the obfuscation area $A_k$ for $k$ shares of precision $\phi_k$ through the intersection of $k$ circles
$c_1 \ldots c_k$, and $M_u$:
\begin{equation}
A_k = \mathrm{area}(M_u \cap c_0 \cap c_{1} \cap \ldots \cap c_{k}) = \pi \ast r_k^2
\end{equation}
Before share generation, the user has to select a map representation $M_u$, which defines the map regions
where the user can possibly be located. $M_u$ is individual for each user,
since different users can be possibly located in different map regions. 

The share fusion algorithm of CSPS is shown in Algorithm~\ref{alg1} (cf. Figure~\ref{fig:rad_inc}a).
It is similar to the OSPS share fusion (Algorithm~\ref{alg1x}). The main difference is that $c_0$
as well as each $c_i$ are intersected with the map representation $M_u$ (lines~2,~7). Note that the radii
are not pre-defined as in OSPS, but each obfuscation has its individual radius defined by the
map-aware share generation algorithm.

\begin{algorithm}
\small
\caption{CSPS: fusion of shares}
\label{alg1}
\begin{algorithmic}[1]
\STATE \textbf{function} $fuse\_k\_shares\_CSPS(M_u, n, c_0, \vec{s_1} \ldots \vec{s_k},  r_1 \ldots r_k)$
\STATE $A_k \gets M_u \cap c_0$ 
\STATE $\vec{p} \gets \vec{p_0}$
\FOR {$i = 1$ \TO $k$}
\STATE $\vec{p} \gets \vec{p} + \vec{s_i}$
\STATE $c_i \gets \{\vec{p}, r_i\}$
\STATE $A_k \gets A_k \cap c_i$
\ENDFOR
\STATE \textbf{return} $A_k$
\end{algorithmic}
\normalsize
\end{algorithm}


Our goal is to preserve the required privacy level by providing
an obfuscation area of a given size. To this end, we adapt the radius such that the intersection
area of $M_u$ and obfuscation circles $c_0 \ldots c_k$ is not smaller than this size.

Algorithm~\ref{alg2} shows the CSPS share generation algorithm with map knowledge (cf. Figure~\ref{fig:rad_inc}b).
In Algorithm~\ref{alg2}, the radius $r_k$ is increased until the area of
$M_u \cap (c_0 \cap c_1 \cap \ldots \cap c_k)$ is equal or greater than
the value of the non-intersected area of $c_k$.
By applying this condition, we ensure that the precision difference between areas $A-k$ and $A_{k+1}$
corresponds to the required $\Delta\phi$.

In lines~4-6, the radius $r_0$ of the initial circle $c_0$ is increased considering the map
representation $M_u$, in order to adjust the size of $A_0 = M_u \cap c_0$.
Also, to adjust the radii of shares $\vec{s_1} \ldots \vec{s_{n-1}}$, $M_u$ is included
in the condition, which defines the target area size (lines~11-13). The remaining steps of the share generation algorithm for CSPS are similar
to Algorithm~\ref{alg2x}.

\begin{algorithm}
\small
\caption{CSPS: generation of shares}
\label{alg2}
\begin{algorithmic}[1]

\STATE \textbf{function} $gen\_n\_shares\_CSPS(n, M_u, r_0, \pi)$
\STATE \textbf{select randomly} $p_0$ \textbf{with} $distance(p_0, \pi) \leq r_0$
\STATE $A_0 \gets \mathrm{area}(c_0)$
\WHILE {$\mathrm{area}(M_u \cap c_0) < A_0$}
\STATE $r_0 \gets \textbf{increase}(r_0, p_0, \Delta r)$
\ENDWHILE 
\FOR {$i = 1$ \TO $n-1$}
\STATE $r_i \gets r_0*(n-i)/n$
\STATE \textbf{select rnd.} $\vec{s_i}$ \textbf{with} $\left|\vec{s_i}\right| \leq 2 \ast r_{i-1}$ \textbf{and} $\pi \in c_i$
\STATE $A_i \gets \mathrm{area}(c_i)$
\WHILE {$\mathrm{area}(M_u \cap \cap_{j=1}^{i}{(c_j)}) < A_i$}
\STATE $r_i \gets \textbf{increase}(r_i, p_i, \Delta r)$
\ENDWHILE
\ENDFOR
\STATE $\vec{s_n} \gets \pi - (\vec{p_0} + \sum_{i=1}^{n-1}\vec{s_i})$
\STATE \textbf{return} $\vec{s_0} \ldots \vec{s_n}$, $r_0 \ldots r_n$

\end{algorithmic}
\normalsize
\end{algorithm}

\begin{figure}
  \begin{center}
    \includegraphics[width=0.7\textwidth]{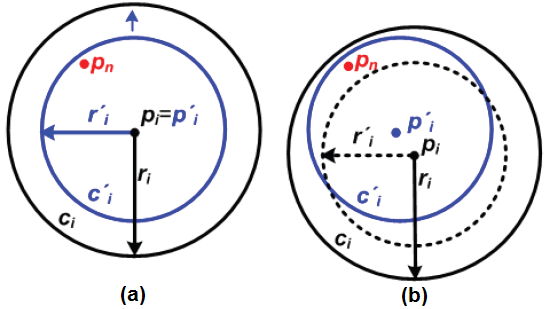}
    \caption{Adjustment of $p_i$ during radius increase: (a) no adjustment of $p_i$;
    (b) randomized adjustment of $p_i$}
    \label{fig:inc_adj}
  \end{center}
\end{figure}

If we used a deterministic algorithm for the area adjustment (lines~5,~12), an attacker could
calculate the inverse function to decrease the size of the obfuscation area.
Therefore, the resulting
center of circle $c_i$ must be randomly shifted together with the radius increase,
so that its original center $c^{\prime}_{i}$ cannot be determined (cf. Figure~\ref{fig:inc_adj}a).

Algorithm~\ref{alg5} shows the $\mathrm{increase}(...)$ function for increasing $r_i$ together with
shifting the center $p_i$ (cf. Figure~\ref{fig:inc_adj}b).
First, for the given $p_i$ we increase the radius $r_i$ until its value provides the required size of the obfuscation area (lines~4-6).
Then $p_i$ is randomly shifted, such that the shift is not greater than
$r_i - r^{\prime}_{i}$ (lines~7-9). Next, we check again whether the current radius $r_i$ still
provides the required size of intersection area (line~10). If this area is
not large enough, we call the function $\mathrm{increase}(r_i,\ldots)$
recursively (line~11). If the intersection area now exceeds the target value $A_i$,
we decrease the current radius $r_i$ until the intersection area hits its limit (lines~12-16).

\begin{algorithm}
\small
\caption{Radius increase with adjustment of $p_i$}
\label{alg5}
\begin{algorithmic}[1]
\STATE \textbf{function} $increase(r_i, p_i, \Delta r)$
\STATE $r^{\prime}_{i} \gets r_i$
\STATE $A_i \gets \mathrm{area}(c^{\prime}_{i})$
\WHILE {$\mathrm{area}(\cap_{j=1}^{i}{(c_j)}) < A_i$}
\STATE $r_i \gets r_i + \Delta r$
\ENDWHILE
\STATE	$x_{shift} \gets \textbf{get\_random\_shift}(p_i, r^{\prime}_{i}, r_i)$
\STATE	$y_{shift} \gets \textbf{get\_random\_shift}(p_i, r^{\prime}_{i}, r_i)$
\STATE $p_i \gets \textbf{shift}(p_i, x_{shift}, y_{shift})$
\IF {$\mathrm{area}(\cap_{j=1}^{i}{(c_j)}) < A_i$}
\STATE $r_i \gets \textbf{increase}(r_i, p_i, \Delta r)$
\ELSE
\WHILE {$\mathrm{area}(\cap_{j=1}^{i}{(c_j)}) > A_i$} 
\STATE $r_i \gets r_i - \Delta r$
\ENDWHILE
\STATE $r_i \gets r_i + \Delta r$
\ENDIF 
\STATE \textbf{return} $p_i, r_i$
\end{algorithmic}
\normalsize
\end{algorithm}

In Figure~\ref{fig:inc_adj}b, we show that after the adjustment of $p_i$, the target position $\pi$ can
be located anywhere within $c_i$; it is not restricted to the obfuscation area which corresponds to the smaller radius $r^{\prime}_{i}$.
Thus, it is not possible for an attacker to reduce the obfuscation area $A_i$
knowing the share generation algorithm and $r^{\prime}_{i}$.

\section{Optimization of Share Placement}

\noindent
So far, we assumed that all servers are equally trustworthy. Now, we consider the case when servers have various trust levels.
Since each LS has an individual trust 
value, the user's position privacy highly depends on the number of selected LSs and the placement of shares on different LSs.
In our previous work \cite{Duerr2011,Skvortsov2012}, we used an equal share placement strategy, i.e., each LS stored shares of the same
precision increase $\Delta_\phi = \Delta^\phi_i$. Now, we want to make sure that an LS with a higher trust 
level can store more precise position information than an LS that has a higher risk of being compromised.


\subsection{Problem Statement: Share Placement under Individual Trust Levels}
\label{sec:pr_st}

\noindent
The problem of share placement on LSs with individual trust levels can be defined as a constrained optimization problem.
The \emph{constraint} is that an attacker cannot derive a position
$\pi_\mathrm{k,attack} = \mathrm{fuse}(s_\mathrm{0},S_k)$ of precision
$\mathrm{prec}(\pi_\mathrm{k,attack}) > \phi_k$ with a probability higher than $P_k(\phi_k)$,
where $S_k$ denotes the set of compromised refinement shares.
That is, the user defines probabilistic guarantees $P_k$ for different precision levels $\phi_k$.

The \emph{optimization} goal is to provide the specified privacy levels and their probabilistic guarantees
by minimizing the number of LSs. 
By minimizing the number of required LSs, we limit the induced overhead of updating (communicating)
and storing shares at multiple servers.


We assume the following values to be given:
\begin{itemize}
	\item master share $s_0$,
	
	\item set $S$ of $n$ refinement shares $\{s_1,\ldots,s_n\}$ to provide the precision (privacy) levels $\phi_k$ for the LBAs,

	\item candidate set $L$ of $m_0$ available LSs, which can store shares $L = \{\mathrm{LS}_1, \ldots, \mathrm{LS}_{m0}\}$,

	\item set of risk values $\{p_1,\ldots,p_{m0}\}$ providing the probabilities for each LS$_i$ of $L$ to be compromised ($p_i \in [0; 1]$),

	\item probability distribution $P_k(\phi_k)$ specifying the required probabilistic guarantees $P_k$ for each precision (privacy) level $\phi_k$.
\end{itemize}

\noindent
Problem: Find 
a share placement $\mathrm{place}(\ldots)$ of $n$ shares to a set $L' \leq L$ of LSs:
\begin{equation}
\mathrm{place}(\{s_1,\ldots,s_n\},L): S \rightarrow L' \subseteq L,
\end{equation}
such that $\left| L' \right|$ is minimal, and the user's security requirements are fulfilled:
\begin{equation}
\label{sh_pl_privacy_restrictions}
\forall~\phi_\mathrm{k,attack} : P_k(\phi_k) > \mathrm{Pr}[\phi_\mathrm{k,attack} \leq \phi_\mathrm{k}]
\end{equation}

\subsection{General Selection \& Placement Algorithm}

\noindent
In our work, we made two assumptions. The first assumption is that increasing the number $m$ of LSs leads
to higher security with regard to probabilistic guarantees of privacy levels.
At the same time, a large $m$ is not desired, since it would increase storage and communication overhead.
Therefore, it is beneficial to incrementally increase $m$ only until
the security requirements are fulfilled. The second assumption is that we can increase security by optimizing
the distribution of shares for a given $m$. For the validation of both assumptions, we refer to our work \cite{Skvortsov15}.

The algorithm for optimal share placement (Algorithm~\ref{alg1pl}) consists of two major steps: (a) selection of a minimum number
$m=|L'|$ of location servers $L'=$ {LS$_1$, LS$_2$, ..., LS$_m$} required to fulfill the given privacy
constraint; (b) optimization of the placement $n$ shares among these $m$ LSs.


The basic idea is to start with the smallest set of LSs and incrementally increase $m$ (lines~2 and~7) until the security
constraints (Equation~\ref{sh_pl_privacy_restrictions}) are fulfilled. In each step, we greedily add the next most trusted LS to set $L'$,
since the subset of the most trusted LSs provides the highest security.
Therefore, the available LSs must be initially sorted by ascending risks $p_i$ (line~3).

For each number of LSs, we first check whether a uniform placement (line~8) where each LS stores an equal number
of shares, independent of its individual risk value, fulfills the user-defined probabilistic guarantees of privacy levels (lines~9-11).
If the non-optimized solution (i.e., uniform placement)
already represents a solution with regard to the constraint, we skip the optimization algorithm
to save energy resources of the mobile device, which executes this algorithm.

\begin{algorithm}
\small
\caption{General Selection \& Placement Algorithm}
\label{alg1pl}
\begin{algorithmic}[1]
\STATE \textbf{function} $place(P_{k}(\phi_k), S, L, m_0, m_{min}, n)$
\STATE $m \gets m_{min} - 1$
\STATE $\mathrm{sort\_by\_ascending\_p_i}(L)$
\STATE $L' \gets \mathrm{get\_selected\_set}(L,m)$
\STATE $solution\_found \gets false$
\REPEAT
\STATE $m \gets m + 1$
\STATE $\mathrm{distribute\_equal}(P_{k}(\phi_k), S, L', m)$
\IF {$\forall~\phi_\mathrm{k} : P_k < P_\mathrm{k,attack}(\phi_\mathrm{k})$}
\STATE $solution\_found \gets true$
\ELSE
\STATE $\mathrm{place\_optimized}(S,L',m)$
\IF {$\forall~\phi_\mathrm{k} : P_k < P_\mathrm{k,attack}(\phi_\mathrm{k})$}
\STATE $solution\_found \gets true$
\ENDIF
\ENDIF
\UNTIL {$(m = m_{0})||(solution\_found)$}
\STATE \textbf{return} $S \rightarrow L'$
\end{algorithmic}
\normalsize
\end{algorithm}


If the uniform (non-optimized) share placement on LSs does not satisfy the user's privacy requirements, we
optimize the placement by relocating shares from less trusted to more trusted LSs (line~12). 
The share placement algorithm invoked in line~12 in the major contribution of this section, and will be presented
in detail in Section~\ref{sec:osp}.


If $m$ reaches the total number of available LSs $m_{0}$, but no solution has been found,
the given security requirements are too hard. Therefore, 
the user should relax the constraints (probabilistic guarantees of privacy levels)
given in Equation~\ref{sh_pl_privacy_restrictions} step by step, 
and execute the algorithm again.

In lines~9 and~13 of Algorithm~\ref{alg1pl}, we determine the probabilistic guarantees of privacy levels of a placement.
We calculate $P_\mathrm{k,attack}$ for different
numbers $k_{attack}$ of compromised LSs for a given placement as follows:
\begin{equation}
P_\mathrm{k,attack}=\sum_{k=k_\mathrm{attack}}^{m}{\sum_{i=0}^{m \choose k}{p_{i,\mathrm{incl}} \cdot p_{i,\mathrm{excl}}}},
\end{equation}
\begin{equation}
p_{i,\mathrm{incl}} = \prod_{j=0}^{m} p_j, \forall p_j \in P_{k,i}
\end{equation}
\begin{equation}
p_{i,\mathrm{excl}} = \prod_{j=0}^{m} ( 1 - p_j ), \forall p_j \notin P_{k,i},
\end{equation}
where $P_{k,i}$ is the set of risks of the $i$th $k$-combination out of $m$ LS risks.

There are ${m \choose k}$ combinations of LS to compromise exactly $k$ LSs. Each
combination has the probability defined by multiplying the risks $p_j$ of (included) $k$ LSs and the
inverse risks $1 - p_j$ of the rest (excluded) $m-k$ LSs. To get a
probability of \emph{exactly} $k$ compromised LSs, we have to sum up the probability of each $k$-combination.
Finally, to get a probability of \emph{at least} $k$ compromised LSs, we sum up the probabilities
corresponding to $\{k, k+1, ..., m\}$ compromised LSs.

\subsection{Optimized Share Placement Algorithm}
\label{sec:osp}

\noindent
In this section, we present an algorithm to solve the optimized share placement problem, which is called from Algorithm~\ref{alg1pl}
through function $\mathrm{place\_optimized}(\ldots)$. Here, we consider the situation when a set $L'$ of
$m = \left| L' \right|$ LSs with lowest risks has been selected and is \emph{fixed}. Thus, an optimized placement of $n$
shares to these LSs in $L'$ must be found, after the uniform placement strategy did not satisfy the user's
security constraints. 
First, we show that this problem is NP-hard. Then, we propose a heuristic solution based on a genetic algorithm.

We selected a strategy that minimizes the risk of revealing the positions of high precision by applying the
principle of \emph{risk balancing}. This principle is well-known from risk-based capital allocation and is commonly
applied if the risk values are available, and there are no correlations between them \cite{Pavlovic08,Albrecht04},
as is the case in our system model, where we do not know any relations between different LS providers.
We call this placement problem the Balanced Risk Placement Problem (BRPP).      

Formally, a share placement $S \rightarrow L' \subseteq L$ has balanced risk if the
proportion of position precisions $\phi_\mathrm{i1,j}$ and $\phi_\mathrm{i2,j}$ stored by
$\mathrm{LS_{i1}}$ and $\mathrm{LS_{i2}}$ respectively ($j = 1 \ldots n$)
is inversely proportional to the corresponding risks $p_{i1}$ and $p_{i2}$ of $\mathrm{LS_{i1}}$ and $\mathrm{LS_{i2}}$:
\begin{equation}
S \rightarrow L' \subseteq L : \forall~i1, i2 \in m, j \in n : \frac{\sum_{j = 1}^{n}{\Delta \phi_{i2,j}}}{\sum_{j = 1}^{n}{\Delta \phi_{i1,j}}} = \frac{p_{i1}}{p_{i2}}
\end{equation}
If the exact equality of proportions is not feasible due to the given risk values and other parameters,
the goal is to find a share placement solution which is close to the best possible solution:
\begin{equation}
\label{sh_pl_problem_form3}
\mathrm{minimize~}max_{i = 1}^{m}{\sum_{j = 1}^{n}{p_{i} \Delta \phi_{i,j}}} - min_{i = 1}^{m}{\sum_{j = 1}^{n}{p_{i} \Delta \phi_{i,j}}},
\end{equation}
under the restrictions of probabilistic guarantees of precision levels given in Equation~\ref{sh_pl_privacy_restrictions}.
This problem 
defines how the function $\mathrm{place\_optimized}(\ldots)$ of Algorithm~\ref{alg1pl} must be implemented.

BRPP is NP-hard, which can be shown by reducing the Agent Bottleneck General Assignment Problem (ABGAP),
which is known to be NP-hard \cite{RePEc88,RePEc1998}, to BRPP. ABGAP is defined as:
\begin{equation}
\mathrm{minimize~}max_{i = 1}^{m}{\sum_{j = 1}^{n}{p_{i} \Delta \phi_{i,j}}}
\end{equation}
ABGAP is equivalent to our placement problem, since one can be polynomially
transformed into another: If we simplify our problem by adding an LS with 0 risk, we can exclude the second
term from Equation~\ref{sh_pl_problem_form3}. This means that in order to solve our problem, we must also solve
ABGAP. Thus, our problem is at least NP-Hard. 

The total number of possible placement combinations for distributing $n$ shares among $m$ LSs is $O(m^n$). 
Since this number grows exponentially with the number of shares, an exhaustive search is very costly for larger $m$ and $n$.
Even relatively small numbers such as $m = 5$ and $n = 15$ require the analysis of more than $3 * 10^{10}$ combinations.
To reduce the computational complexity, we implement a linear-time heuristic.
Our goal is not to find the best placement among all possible combinations, but to find a
placement which satisfies the required probabilistic guarantees.
To this end, we need a strategy that guides our search for a secure placement 
in a reasonable (linear) time.

In general, problem-specific heuristics or meta-heuristics
can be used to find heuristic solution. We applied the
meta-heuristic of \emph{genetic algorithms}~\cite{weicker2002}. 
Genetic algorithms reproduce the process of biological evolution. They work on multiple solution candidates by
combining and mutating
them into new possible solutions. Each new solution (in our case, a share placement)
is rated according to a fitness (objective) function defined by Equation~\ref{sh_pl_problem_form3}.
Then, the best placements in terms of the objective functions are selected, and the cycle can repeat
until the goal is reached or the limit of cycles is achieved.

\begin{algorithm}
\caption{Genetic Algorithm for Share Placement}
\small
\label{alg2pl}
\begin{algorithmic}[1]
\STATE \textbf{function} $place\_optimized(S, L', m)$
	\STATE $t \gets 0$
	\STATE $Popul[1 \ldots 10] \gets RandomPlacement(S,L',m)$
	\WHILE{$t < 200$ \textbf{and} $\forall~P_k < P_\mathrm{k,attack}$} 
		\FOR{$p = 1$ to $40$} 
			\STATE $i_1 \gets RandomInteger(m)$; $i_2 \gets RandomInteger(m)$
			\STATE $u \gets RandomBoolean()$
			\IF{$u$}
				\STATE $PopulTemp[p] \gets \mathrm{Cross}(PopulTemp,i_1,i_2)$
			\ENDIF
			\STATE $PopulTemp[p] \gets \mathrm{Mutate}(PopulTemp[p])$
	\ENDFOR
		\STATE $\mathrm{Evaluate}(PopulTemp)$
	\STATE $Popul \gets \mathrm{Select10Best}(PopulTemp)$
	\STATE $P_{attack}(\phi) \gets \mathrm{BestLevels}(Popul)$
	\STATE $t \gets t + 1$
	\ENDWHILE
\end{algorithmic}
\normalsize
\end{algorithm}

We implemented a genetic algorithm for share placement as shown in Algorithm~\ref{alg2pl}. 
The input parameters are the probabilities $P_{k}(\phi_k)$, the set $L$ of size $m$, and the fixed set of shares $S$ of size $n$.
First, we define the initial population as 10 random placements (line~3). Then, we build a population of 40 new placements
by recombining two placements with a uniform crossover (with a probability of 50\%) (lines~5-10).
Afterwards, the placement is mutated by changing one assignment randomly (line~11), ensuring that theoretically all 
possible placements could be created.

Next, the 40 created placements are rated according to the objective function, and 10 best placements are selected (lines~13-14). 
This cycle is iterated 200 times or stopped if the conditions of Equation~\ref{sh_pl_problem_form3} are satisfied (lines~4-17).
The value of 200 iterations is selected such that it ensures convergence. Our experiments have shown that we already achieve a near-optimal
placement solution after about 20 iterations.
If after all cycles the probabilistic guaranties are still not satisfied (line~4), we say that the solution cannot be found
for the given input parameters.

\section{Optimization of Position Update Algorithm} 

\noindent
Our basic approach \cite{Duerr2011,Skvortsov2012} described in Section~\ref{sec:sys_mod},
allows for sharing user's position among multiple non-trusted LSs,
but for the price of increased communication overhead.
A complete set of shares has to be re-generated and sent to the corresponding set of LSs every time a position update event is triggered.
More precisely, an MO must send $n$ messages
with new position shares to $n$ different LSs, while an LBA must receive $k$ messages from $k$
LSs in order to obtain the position of the $k$th precision level. 
This principle can produce a high communication overhead, e.g., if the update rate is high and the number of LSs is large.
However, in many cases such as sporadic movements of the MO, the re-generation and update of the whole share set causes redundancy.
Hence, our goal is to send a smaller number of messages than $n$ after each position change.

In this section, we begin by defining the problem of message reduction. Then, we describe an optimized position
update algorithm for position sharing approach. 

\subsection{Problem Statement: Minimization of Position Update Messages}

\noindent
We formulate the reduction of position updates as a constrained optimization problem.

The \emph{optimization} goal is to reduce the total number of position update messages being sent
from MOs to LSs (denoted as the number of messages $N^{MO-LS}$). 
The \emph{constraints} are that there should be no change of position
precision $\phi_k$ as a result, as well as no reduction of the user's probabilistic guarantees $P_{k,attack}(\phi_k)$ of precision (privacy) levels.

We assume the following values to be given:
\begin{itemize}
\item $n$ location servers,
\item the MO's previous consecutive precise position $\pi_{i}$, i.e., the position before $\pi_{i+1}$
(the algorithm is run on the MO side, which means that the MO's own precise positions are available),
\item the MO's next consecutive precise position $\pi_{i+1}$, i.e., the position after $\pi_{i}$,
\item master share $s_0$ generated for $\pi_{i}$,
\item set $S^{i}$ of $n$ refinement shares ${s_1 \ldots s_n}$ generated for $\pi_{i}$,
\item probability distribution $P_k(\phi_k)$, which specifies the required probabilistic guarantees
for each precision level $\phi_k$.
\end{itemize}

Problem: Find the set of shares $S^{i+1}_{opt}$
such that $S^{i+1}_{opt}$ requires the minimal number of update messages from MOs to LSs.
In other words, $S^{i+1}$ ($S^{i+1}=S^{i+1}_{opt}$) and $S^i$ should differ in as few shares as possible, i.e., in $S^{i+1}$
as many shares as possible should be reused from $S^i$.

The concatenation of all shift vectors of $S^{i+1}_{opt}$ must point
to $\pi_{i+1}$: 

\begin{equation}
\label{eq:locup_opt1a}
S^{i+1}_{opt} = \{s^{i+1}_{0} \ldots s^{i+1}_{n}\} ~~:~~\sum^{n}_{k=0}{s^{i+1}_{k}} = \pi_{i+1}, 
\end{equation}

The precision $\phi_k$ of each imprecise position $p^{i+1}_{k}$ derived by share fusion after obtaining
the minimized set $S^{i+1}_{opt}$ has to be the same as
the precision of the corresponding imprecise position $p^{i}_{k}$ derived from the original set of shares $S^{i}$:

\begin{equation}
\label{eq:locup_opt2}
\forall S^{i+1}_{k} \in S^{i+1}_{opt},~~S^{i}_{k} \in S^{i}~~:~~\phi_{k}(p^{i+1}_{k}(S^{i+1}_{k})) = \phi_{k}(p^{i}_{k}(S^{i}_{k}))
\end{equation}

Finally, the set of shares $S^{i+1}_{opt}$ must also satisfy
the current user's privacy requirements, i.e., each further $k$th share must provide the pre-defined probability $P_{k}(\phi_k)$:

\begin{equation}
\label{eq:locup_opt2a}
\forall~\phi_\mathrm{k,attack}~~:~~P_k(\phi_k) > {Pr}[\phi_\mathrm{k,attack} \leq \phi_\mathrm{k}];
\end{equation}


Note that we do not assume that an MO's complete trajectory is available. We consider only the neighboring consecutive position
updates. 
Hence, we cannot apply statistical analysis of the past positions and the respective parameters such as speed, and therefore we do not consider
approaches for preserving privacy of a complete trajectory.

\subsection{Optimized Position Update Algorithm}

\noindent
Depending on the movement scenario, different position update approaches can be beneficial.

The key factor which separates sporadic and continuous update scenarios
is the relation of the distance traveled between two consecutive updates
and the radii of obfuscation circles.

\begin{figure}
\begin{center}
\includegraphics[width=0.6\textwidth]{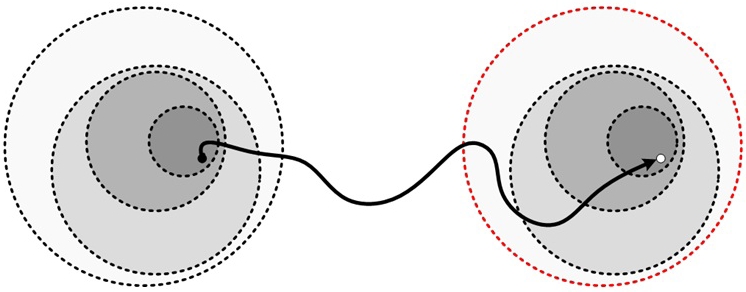}
\caption{Large movement of MO; two consecutive master shares do not intersect}
\label{fig:locup_case2}
\end{center}
\end{figure}

If MO moves fast or the update rate is very low, the new MO's master share can be located completely
outside the previous master share, as depicted in Figure~\ref{fig:locup_case2}. 
The condition of having no intersection between two consecutive master shares is: 
\begin{equation}
\label{eq:locup_case2_cond2}
\mathrm{distance}(p_{i+1}^0, p_i^0) > 2*r_0
\end{equation}

In the following, we will only consider sporadic position updates with long time intervals between updates rather
than the continuous tracking of users. Thus, we assume that the time between updates is long enough
such that there is no relation between consecutive updates, which an attacker could exploit.
An optimized position sharing approach focusing on continuous updates using, for instance,
dead-reckoning techniques, has been described by us in~\cite{Riaz2015}.

The main idea of our optimized position update algorithm is that under the condition of Equation~\ref{eq:locup_case2_cond2},
we can recalculate and update only the master share while keeping the refinement shares unchanged.

Now we can estimate the communication costs for the optimized position update approach. Since only one share has to be updated,
the number of sent messages between MO and LS is $N^{MO-LS}_{\mathrm{opt}} = 1$.
In contrast, the basic position update approach requires
all shares to be re-generated and sent, i.e., $N_{\mathrm{basic}} = n$.
The resulting reduction rate of the communication cost is:
\begin{equation}
\label{eq:locup_case2:saving_mo_ls}
R′^{MO-LS}_{\mathrm{opt}} = \frac{n - 1}{n}
\end{equation} 

Note that since $N^{MO-LS}_{\mathrm{opt}} = 1$, our optimized position updated algorithm is always optimal
w.r.t. the cost of communication between MO and LSs.

The pseudocode for the optimized position update approach run by MO is presented in Algorithm~\ref{alg:locup_lua}.
Before sending a location update, the MO determines whether the optimized approach is applicable
by checking the condition of Equation~\ref{eq:locup_case2_cond2} (line~2).
If the condition of Equation~\ref{eq:locup_case2_cond2} is satisfied, only a new master share
has to be generated and sent (lines~3-4) to the corresponding LS.
The refinement shares $s_1 \ldots s_n$ will remain the same without causing any inconsistency during their fusion.
This is enabled by the fact that shares are relative shift vectors, while the absolute coordinates are only contained
in the master share $s_0$.
If the condition of Equation~\ref{eq:locup_case2_cond2} is not satisfied, the basic position update protocol is applied (lines~5-7).
\begin{algorithm}
\small
\caption{Optimized Position Update Algorithm}
\label{alg:locup_lua}
\begin{algorithmic}[1]
\STATE \textbf{function} $update\_shares(\vec{\pi_{i}}, \vec{\pi_{i+1}}, n, \vec{s_0} \ldots \vec{s_n})$
\IF {$\mathrm{distance}(\vec{p_{i}^0}, \vec{p_{i+1}^0}) > 2*r_0$}
\STATE $\mathrm{update\_shares\_opt}(\vec{\pi_{i}}, \vec{\pi_{i+1}}, n, \vec{s_0})$
\STATE $\mathrm{send}(\vec{s_0})$
\ELSE
\STATE $\vec{s_{0}} \ldots \vec{s_n} \gets regenerate\_all\_shares(\vec{p_{i+1}}, n, \phi_{min}, \Delta_\phi)$
\STATE $\mathrm{send}(\vec{s_0} \ldots \vec{s_n})$
\ENDIF
\end{algorithmic}
\normalsize
\end{algorithm}

\subsection{Security of Position Updates}

\noindent The \emph{first privacy requirement} corresponding to our
problem statement (Equation~\ref{eq:locup_opt2}) is that the position update optimization must not
reduce the obfuscation area, i.e., it must not cause an undesired increase in position precision $\phi_k$.
Regarding this condition, we can state that the proposed optimized location
update algorithm does not reduce the number of shares, i.e., it does not change the precision level available to authorized LBAs.
In other words, the smaller number of shares sent from MO to LS does not affect the number of shares provided to LBAs.
Therefore, no change of precision occurs.

The \emph{second privacy requirement} corresponding to our problem statement
(Equation~\ref{eq:locup_opt2a}) is the probabilistic metric $P_k(\phi_k)$.
According to the optimized location update algorithm, separate location updates only update master shares $s^i_0$.
The remaining refinement shares $s^i_1 \ldots s^i_n$ and the corresponding refined obfuscation circles remain unchanged, i.e.,
the attacker does not get new knowledge.
Thus, the randomness of share generation is preserved by the unchanged share generation algorithms (Algorithm~\ref{alg2x}, Algorithm~\ref{alg2}),
and the probabilities $P_k(\phi_k)$ are the same as in the basic position sharing approaches.

\section{Evaluation}

\noindent
Next, we present the evaluation of our share placement algorithm and the optimized position update algorithm.
We start with an evaluation of the share placement runtime performance, before we compare the probabilistic guarantees
of precision levels of our basic approach with the probabilistic guarantees resulting from optimized share placement.
Then, we evaluate the communication cost of the optimized position update algorithm.

\subsection{Runtime Performance of Placement Optimization}

\noindent
According to the principle of position sharing, share placement has to be calculated on the mobile
device of the user, since it is the only trusted entity in our system model. Since mobile devices
are typically restricted in terms of processing power and energy, the runtime of our share placement
algorithm is crucial. Therefore, we measured the runtime of calculating the placement of a set of shares on a state of the art
mobile device (HTC Desire with Android OS, CPU: 1 GHz Qualcomm QSD8250 Snapdragon, memory: 576 MB RAM). 
We tested the full number of cycles of the genetic algorithm, without terminating the algorithm under
the ``solution found'' condition (i.e., we have tested the worst case scenario, when the solution is not feasible
for the given parameters). The number of LSs was given as $m = 5; 10; 20$, and the number of shares $n$ is in the interval $[m; 50]$.

\begin{figure}
  \begin{center}
    \includegraphics[width=1.0\textwidth]{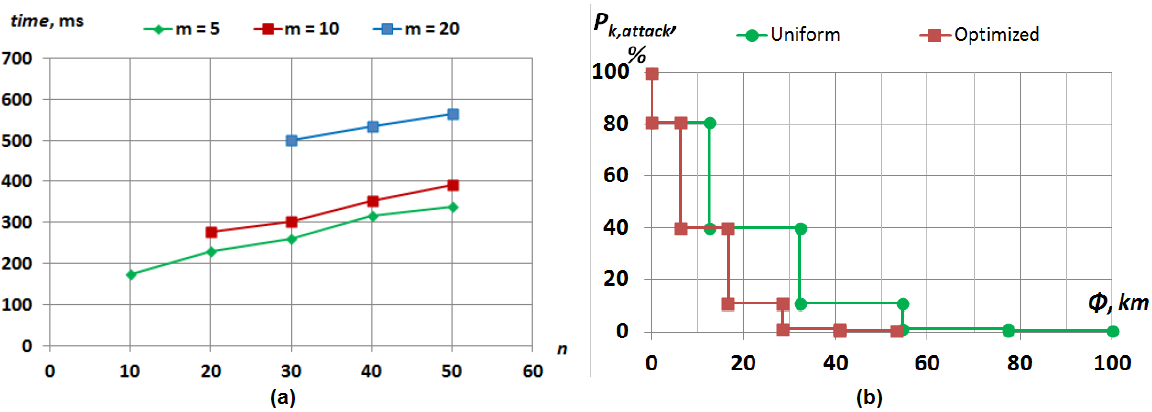}
    \caption{(a) Computational cost of genetic share placement algorithm;
		(b) Improvement of probabilistic guarantees through placement optimization}
    \label{fig:sh_pl_eval12}
  \end{center}
\end{figure}

Figure~\ref{fig:sh_pl_eval12}a shows the average runtime for placing $n$ shares on $m$ LSs.
As our evaluation shows, the proposed Algorithm~\ref{alg2pl} has linear complexity 
and is executed in less than one second even for larger input parameters ($m = 20$, $n = 50$).
Therefore, we conclude that the algorithm is suited also for resource-poor mobile devices.

\subsection{Probabilistic Guarantees of Precision Levels after Placement Optimization}

\noindent
Next, we compare the resulting probabilistic guarantees of precision levels of optimized share placement compared to a basic (non-optimized)
placement algorithm. We placed $n = 15$ shares on $m = 5$ LSs with heterogeneous risks. The values of risk were chosen uniform at random
from the interval $[0;0.5]$: $p_1 = 0.4932$; $p_2 = 0.3292; p_3 = 0.2344; p_4 = 0.1788; p_5 = 0.0925$.
The basic algorithm distributes an equal number of shares (3) to each LS, while the optimized placement
placed 1, 2, 2, 3, and 7 shares onto the given LSs.

Figure~\ref{fig:sh_pl_eval12}b depicts the probabilities $P_\mathrm{k,attack}$ for the different precision levels $\phi$. Note that the precision levels $\phi_\mathrm{k,attack}$ which correspond to the
probability levels $P_\mathrm{k,attack}$
are calculated as the weighted average of position precisions 
of each possible $k$-combination.
In Figure~\ref{fig:sh_pl_eval12}b, we can see that for the same precision $\phi$, the probabilities
$P_\mathrm{k,attack}$ are significantly lower for the most $\phi$ values.
In other words, the optimized placement of shares has shifted the peaks of disclosure probability to the left.
For example, for $\phi = 25$ km, $P_\mathrm{k,attack} = 10.7\%$ for the optimized share placement
and $P_\mathrm{k,attack} = 40.1\%$ for the uniform share placement.

\subsection{Communication Cost after Position Update Optimization}

\noindent
Next, we evaluate the communication overhead of the basic and optimized location update protocol.
As performance metric we use the number of messages sent to LSs by the different update protocols. 

For this evaluation, we used the GeoLife data set \cite{ZhengXM10}.
including real trajectories of periodic trips to work, or hiking and biking trips. From this data set,
we have selected two trips with different update intervals and distances between updates. The first trip
has shorter update intervals (up to 15 s) and distances between updates (tens of meters). The second trip has
longer update intervals (several minutes to more than one hour) and distances between updates (up to kilometers). 

\begin{figure}
\begin{center}
\includegraphics[width=0.55\textwidth]{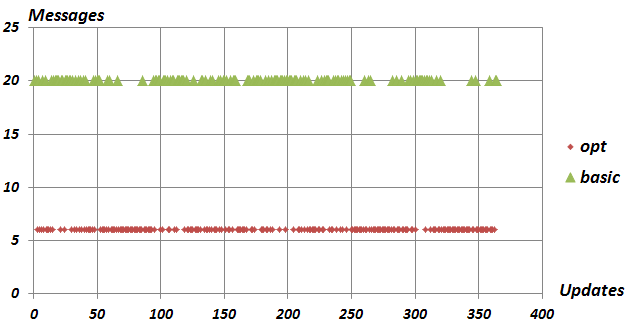}
\caption{Location updates with $r_0 = 5$ m; $n = 5$} 
\label{fig:locup_eval1a}
\end{center}
\end{figure}

\begin{figure}
\begin{center}
\includegraphics[width=0.95\textwidth]{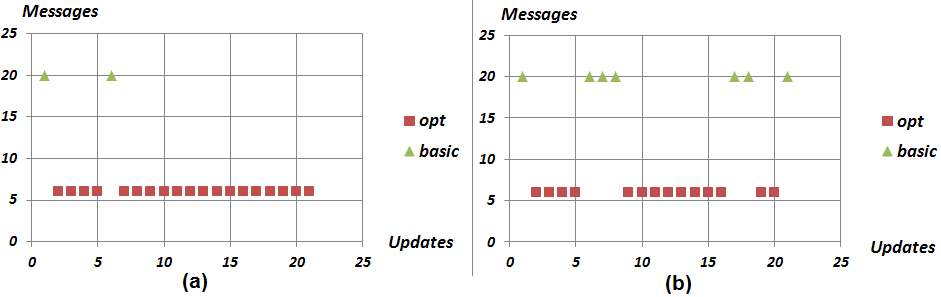}
\caption{Location updates with $n = 5$: (a) $r_0 = 50$ m; (b) $r_0 = 100$ m}
\label{fig:locup_eval2ab}
\end{center}
\end{figure}

Figure~\ref{fig:locup_eval1a} shows the number of update messages sent for each location update (i.e., per position change of the trip)
by the basic and optimized update protocol, respectively, for short update intervals. The number of shares is $n=5$.
The radius of the master share is $r_0=5$~m (Figure~\ref{fig:locup_eval1a}), $r_0=50$~m and $r_0=100$~m (Figures~\ref{fig:locup_eval2ab}a,\ref{fig:locup_eval2ab}b).
The optimized update approach is often selected in Figure~\ref{fig:locup_eval1a} due too small radius of the master share
(generating 6 messages per update), while for a larger radius and shorter update intervals the basic approach is applied
more often (generating 20 messages per update).
For the given parameters, the basic approach generates 7280 messages in total for the given 364 position updates.
The optimized approach generates 4158 messages in total (sending 6 messages instead of 20 in 223 out of 364 position updates),
which corresponds to a reduction rate of 42.8\%.

Figure~\ref{fig:locup_eval2ab} shows the number of update messages per location update for the trip with long update
intervals. Again the number of shares is $n=5$.
In Figure~\ref{fig:locup_eval2ab}a and Figure~\ref{fig:locup_eval2ab}b, we set $r_0=50$ m and $r_0=100$ m, respectively.
The basic approach generates 420 messages in total, with 20 messages per each update.
The optimized position update algorithm with 6 messages per update can be selected often since the distance between the updates is large.
The non-optimized updating with 20 messages per update is selected more often with $r_0=100$ due to a larger number of master share intersections.
As a result, the optimized position update algorithm generates 154 for $r_0=50$ and 224
messages in total for $r_0=100$ (reduction rate of 63.3\% and 46.7\% respectively).

\begin{figure}
\begin{center}
\includegraphics[width=0.6\textwidth]{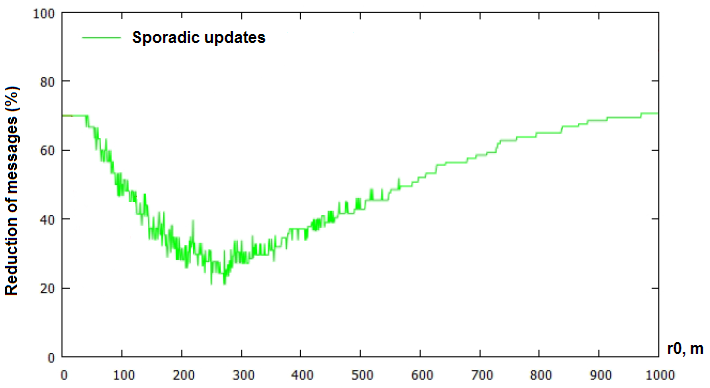}
\caption{Saving ratio (reduction of update messages, \%) depending on radius $r_0$; $n = 5$} 
\label{fig:eval_r0_sporadic_only}
\end{center}
\end{figure}

In Figure~\ref{fig:eval_r0_sporadic_only}, we show the relative communication overhead of the optimized update
protocol in relation to the basic update protocol for different radii $r_0$.
Here, we used the trip with long update intervals.
We can see that a larger $r_0$ usually causes fewer updates for larger $r_0$.
At the same time, for smaller $r_0$ values (for the given mixed data set: between 0 and 300 m),
a smaller $r_0$ leads to fewer intersections and, therefore, to a smaller number of update messages.

\section{Discussion}

Finally, we discuss limitations and possible extensions to our approach presented in this paper.

In this paper, we have only considered sporadic updates of positions rather than continuous updates.
This is a reasonable assumption for many systems based on user check-ins to locations. Our basic
assumption for sporadic updates is that there is no relation between two consecutive updates. In order
for this assumption to hold, there must be sufficient time between consecutive updates depending on the
distance between updates since otherwise an attacker can exclude some areas from the obfuscated area to
increase the precision of obfuscated positions.
In particular, we need to consider the so-called
Maximum Velocity Attack \cite{ghinita2009}.
This attack limits the obfuscation area of subsequent updates by assuming a maximum MO speed and
distance travelled since the last update.
In order not to be prone to such attacks with our approach, the time interval between consecutive updates
must be long enough such that the user could travel between both positions at least two times.
Otherwise, we need to suppress the update.
This is the well-known counter-measure against such an attack, described 
in the literature \cite{ghinita2009,Wernke2013}.

It is also possible to consider a probability distribution (pdf) based on the movement correlation
instead of one based on intersection of the binary movement
boundary. However, such an approach requires analysis of the trajectory correlation pattern
and is beyond the scope of this paper. We refer to our work \cite{Riaz2015},
which extends the position sharing approach by considering trajectory data, and
which estimates the resulting probabilistic privacy guarantees.


\section{Conclusion}

\noindent
In this paper, we described our position sharing approach, which distributes position information of a mobile user
among multiple location servers of non-trusted providers in the form of separate position shares.
This approach has several interesting properties like no single point of failure with respect to privacy
and graceful degradation of privacy with the number of compromised location servers.

In this work, we have further improved our basic position sharing approach.
The first extension increases the user's location privacy if the available location servers have different trust levels.
Based on a probabilistic security metric, we 
proposed an approach
which improves the user security by selecting the minimal required
number of location servers and by optimizing the distribution of position shares among these servers.
Our solution has linear complexity and in practice can be executed 
on resource-poor mobile devices.
Our evaluations show that that this optimized share placement provides a
30\%--40\% lower probability of a user being discovered by an attacker at the same precision (privacy) levels.

The second extension of our basic approach is the optimization of the communication cost caused by
position updates. 
Our calculation of communication cost takes into account the messages sent from mobile users to location servers
as well as the messages sent from location servers to location-based applications.
We have shown that our position update approach
significantly reduces the communication cost achieving a traffic reduction of up to 60\% of the non-optimized protocol version.


\section*{Acknowledgements}
We gratefully acknowledge the German Research Foundation (Deutsche Forschungsgemeinschaft, DFG)
for financial support of this research (PriLoc project).

\section*{References}
\newcommand{\BIBdecl}{\setlength{\itemsep}{0.25 em}}
\bibliographystyle{IEEEtran}
\footnotesize
\bibliography{bib}

\end{document}